\documentclass[
    aip
    , jcp
    , preprint
    , final         
    , amssymb, amsmath
]{revtex4-1} 

\usepackage[utf8]{inputenc}
\usepackage[T1]{fontenc}

\usepackage[english]{babel}
\usepackage{lmodern}
\usepackage[babel=true]{microtype}

\usepackage{graphicx}
\usepackage{booktabs}
\usepackage[caption=false]{subfig}
\usepackage[dvipsnames]{xcolor}
\definecolor{lgray}{rgb}{.75,.75,.75}
\definecolor{cardinal}{rgb}{0.77, 0.12, 0.23}
\definecolor{prettyblue}{RGB}{0, 157, 209}
\definecolor{markblue}{RGB}{0, 76, 153}

\newcommand{\refa}{(\textbf{a})}
\newcommand{\refb}{(\textbf{b})}
\newcommand{\refc}{(\textbf{c})}
\newcommand{\refd}{(\textbf{d})}

\newcommand{\eg}{e.\,g.~} 
\newcommand{\ie}{i.\,e.~} 
\newcommand{\etal}{\textit{et\,al.} } 

\newcommand{\grad}{^\circ_{}}
\newcommand{\pinit}{\rho}
\newcommand{\pfree}{\rho_{\mathrm{free}}^{}}

\newcommand{\orderparticle}{\Theta_{p}^{}}

\newcommand{\ordersystem}{\Theta_{s}^{}}

\newcommand{\Nmin}{N_{\mathrm{c}}\geq30}
\newcommand{\pairprobability}{P_\mathrm{c}(r,t_0+\Delta t)}
\renewcommand{\exp}[1]{\mathrm{e}^{#1}} 

\begin{document}
\noindent The following article has been accepted by the Journal of Chemical Physics. After it is published, it will be found at \url{http://dx.doi.org/10.1063/1.5086618}.
\title{Non-equilibrium effects of micelle formation as studied by a minimum particle-based model}
\date{April 15, 2019}
\author{Simon Raschke}\email{\url{simon.raschke@uni-muenster.de}}
\author{Andreas Heuer}\email{\url{andheuer@uni-muenster.de}}
\affiliation{Westfälische Wilhelms-Universität Münster, Institut für physikalische Chemie, Corrensstraße 28/30, 48149 Münster, Germany}

\begin{abstract}
    The formation of self assembled structures such as micelles has been intensively studied and is well understood. %
    The ability of a solution of amphiphilic molecules to develop micelles is depending on the concentration and characterized by the critical micelle concentration (cmc), above which micelle formation does occur. %
    Recent studies use a lattice approach in order to determine cmc and show that the correct modelling and analysis of cluster formations is highly non-trivial. %
    We developed a minimalistic coarse grained model for amphiphilic molecules in the continuum and simulated the time evolution via dynamic Monte Carlo simulations in the canonical (NVT) ensemble. %
    Starting from a homogeneous system we observed and characterized how the initial fluctuations, yielding small aggregates of amphiphilic molecules, end up in the growth of complete micelles. %
    Our model is sufficiently versatile to account for different structures of surfactant systems such as membranes, micelles of variable radius and tubes at high particle densities by adjusting particle density and potential properties. %
    Particle densities and micellization rates are investigated and an order parameter is introduced, so that the dependence of the micellization process on temperature and surfactant density can be studied. %
    The constant density of free particles for concentrations above cmc, \eg as expected from theoretical considerations, can be reproduced when choosing a careful definition of free volumes. %
    In the cmc regime at low temperatures different non-equilibrium effects are reported, occurring even for very long time-scales. %
\end{abstract}

\maketitle

\section{Introduction}
The aggregation process of micelle and vesicle forming surfactant systems has been intensively studied for decades and the basic physical mechanisms are well understood\cite{self_theory,cmc-sds}. %
Furthermore, a lot of atomistic and coarse grained simulations regarding membrane formation of lipids and surfactants have been performed~\cite{smit1993computer,tarek1998molecular,goetz1998computer,marrink2000molecular,shelley2000computer,hakobyan2013phase}. %
Minimalistic simulation models in continuous space, taking the coarse graining down to three beads per molecule have been analyzed\cite{cooke2005solvent}. %
Beyond continuous models also a very efficient square and cubic lattice model has been devised to model the process of micellization\cite{larson1985monte,bernardes1994monte,mackie1997aggregation,floriano1999micellization,lisal2003formation}. %
The insight of this work is of interest for the understanding of the cell membrane functionality in biological research\cite{zhang2003fabrication,pohorille2009self}. %
Other materials, such as single-wall carbon nanotubes\cite{angelikopoulos2009dispersing}, which offer a nucleation site, were studied as well and show related aggregation behavior. %

\begin{figure}[b]
    \includegraphics{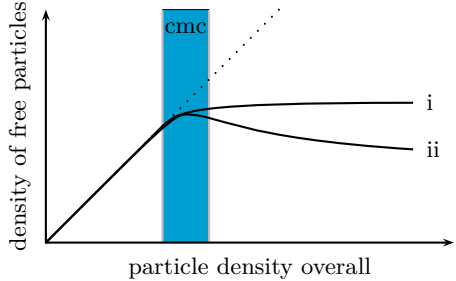}
    \caption{Schematic description of the cmc domain for the density of free particles as a function of the overall particle density (i) with and (ii) without inaccessible volume calculation.}\label{fig:cmc_scheme}
\end{figure}

A key property to characterize the formation of micelles, is the critical micelle concentration (cmc). %
Experimentally, the cmc can be obtained from UV-absorption spectroscopy, fluorescence spectroscopy and electrical conductivity\cite{cmc-surfactant}. %
In theoretical studies the cmc is typically derived from determination of the free or the oligomeric surfactant concentration in the system\cite{lazaridis2005implicit,vishnyakov2013prediction}. %
Ideally, one would obtain the dependence of the free surfactant concentration on the total concentration as sketched in figure~\ref{fig:cmc_scheme}. %
In practice this ideal behavior is hampered by different effects. %
(1)~Typically this curve displays a maximum. %
As discussed by \citet{cmc-surfactant2} this effect is related to the increase of the inaccessible volume with increasing number of micelles, \ie for larger surfactant concentration (see below for a closer discussion). %
The effect is well known in literature and occurs in lattice\cite{mackie1997aggregation,talsania1997monte} and continuous\cite{von1997stochastic} model simulations. %
(2)~In particular for higher temperatures one observes a broad crossover. %
As a consequence, different definitions of the cmc (\eg~based on (1) or based on the properties of the osmotic pressure) start to deviate significantly. %
(3)~Close to the cmc at lower temperatures the system requires a long time to equilibrate. %
This behavior of long equilibration times at low concentrations, where cmc usually resides, is a well known challenge, due to much longer cpu times for non coarse grained force fields\cite{velinova2011sphere,perez2013modeling}. %
It becomes even more pronounced, the more rugged the potential energy landscape becomes, \eg in all atom molecular dynamics simulations\cite{kraft2012modeling}. %
Thus, for a given computer time scale there is always a temperature below which no equilibration is possible. %
(4)~Significant fluctuations are observed, hampering a precise determination of the relevant physical observables\cite{cmc-surfactant2}. %

The key goal of the present work is to shed new light on the properties of micelle formation close to cmc based on computer simulations of surfactant molecules in continuous space. %
Due to the long time scales, anticipated for these simulations, it is essential to use simple model systems. %
Here, we present a new minimalist model system, giving rise to the formation of spherical micelles. %
It consists of small rod-like particles interacting by a modified Lennard-Jones potential. %
We choose a stochastic dynamics by using simple Monte Carlo moves. %
We take care that the potential energy landscape of two surfactants is sufficiently smooth to avoid the trapping in local energy minima. %
This model allows us to reach sufficiently long simulation times and thus to study the nature of the fluctuations close to cmc in detail, including the characterization of significant non-equilibrium effects. %
Inspired by Ref.~\cite{cmc-surfactant2} we explicitly take care of the excluded volume in order to get rid of the significant maximum of the free surfactant concentration as a function of the overall surfactant concentration. %
In particular we will show that the formation of micelles contains processes which take place on exponentially long time scales.

\section{Method}
\subsection{Potential}\label{sec:potential}

\begin{figure}
    \includegraphics{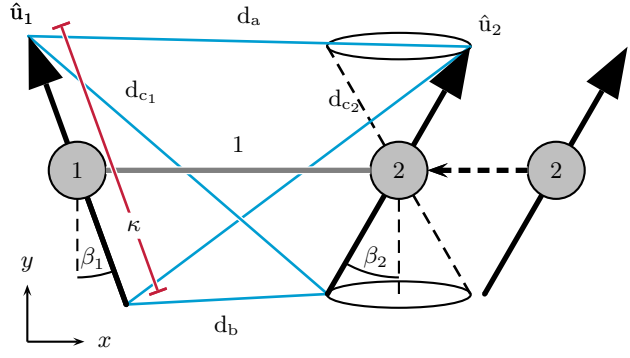}
    \caption{Schematic description of two interacting molecules with orientation angles \( \beta^{}_i \) and length $\kappa$. %
    The $\hat{u}_i$ denote the respective unit vectors along the molecules. %
    The connection vector and $\hat{u}_1$ span a plane, where the connection vector denotes the x direction and the y direction is orthogonal to x in this plane. $\beta^{}_1$ and $\beta^{}_2$ are then defined as the angle between the y-axis and $\hat{u}_i$.
    Furthermore, the different distances between the heads and tails are indicated. %
    For their definition the second molecule is shifted in a parallel manner from its original position to the equilibrium distance. %
    See the main text for more details. %
    }\label{fig:particles_scheme}
\end{figure}

\begin{figure}
    \includegraphics{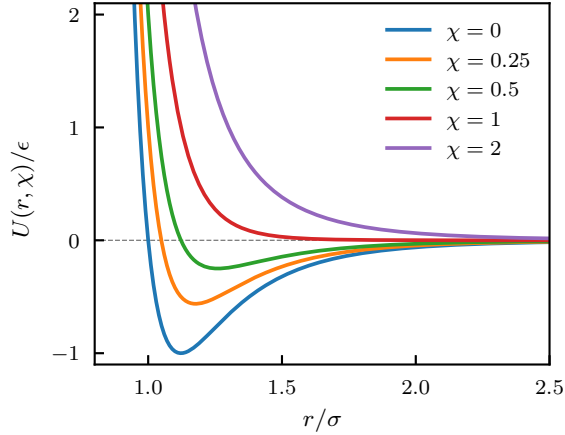}
    \caption{Potential energy of the modified Lennard-Jones potential (eq.~\ref{eq:modLJ}) as a function of the particle distance \(r\) and the magnitude of the orientation dependent term \( \chi \).}\label{fig:potprop}
\end{figure}

\begin{figure*}
    \begin{tabular}{cc}\vspace*{-3ex}
        \subfloat[\label{sfig:landpp}]{ \includegraphics{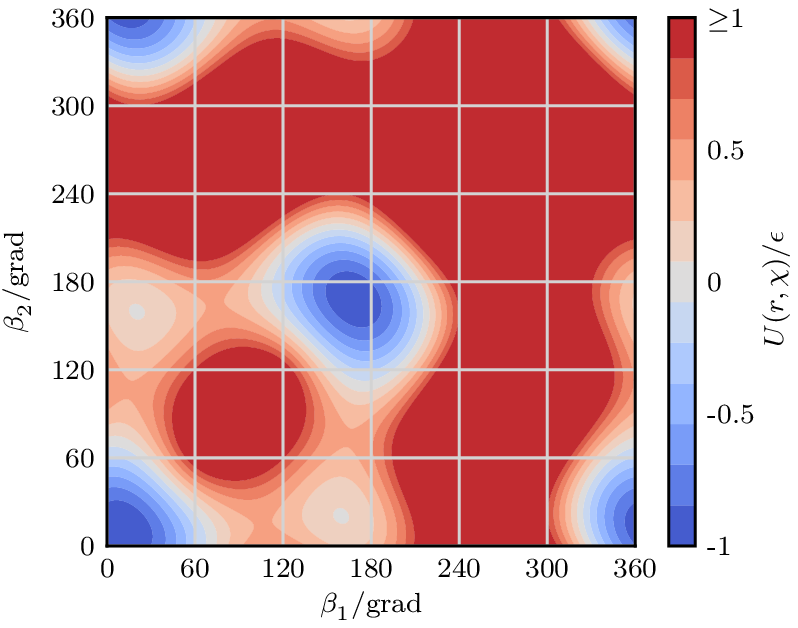} } &
        \subfloat[\label{sfig:landbr}]{ \includegraphics{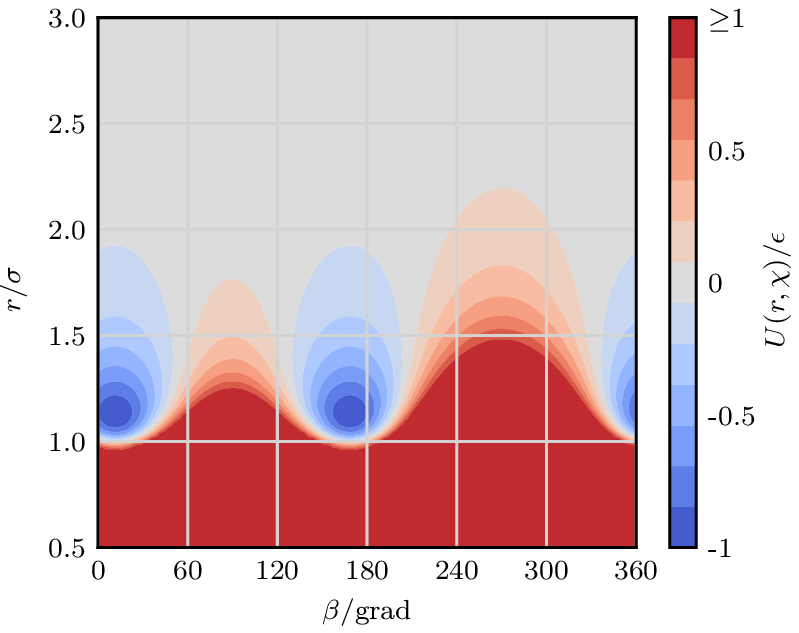} } \\
        \subfloat[]{ \includegraphics{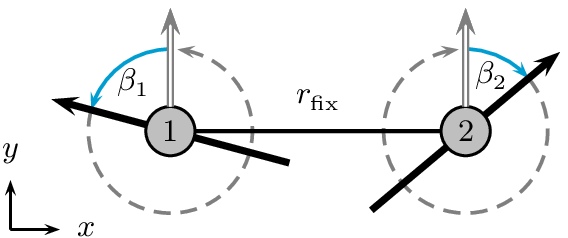} } &
        \subfloat[]{ \includegraphics{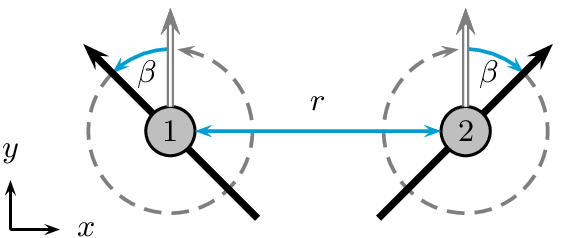}   } \\
    \end{tabular}
    \caption{Potential energy landscape of two particles with the optimum angle \( \gamma=11\grad \) \refa~rotating independently with \( \beta^{}_{1,2} \) in the x,y-plane in the Lennard-Jones minimum distance \( r^{}_\mathrm{fix}=1.122\,\sigma \) and \refb~rotating
    simultaneously with \(\beta\) in the x,y-plane at various distances. %
    The changing parameters used to generate the energy landscapes are depicted schematically with blue arrows at \refc~corresponding to the \( \beta^{}_{1,2} \) landscape and \refd~corresponding to the \(\beta, r\) landscape. }\label{fig:land}
\end{figure*}

Here we present a minimalist model with which we can study micellization behavior on long time scales. %
We combine a Lennard-Jones potential for the distance-dependent interaction with an appropriately chosen angular-dependent term. %
Examples for such an approach can be found in literature \citep{gay_berne,fejer2011self}. %
A surfactant molecule is chosen as a point like particle with an additional vector-type degree of freedom of length \(\kappa\), expressing its orientation. %
The pair-wise interaction of two molecules is chosen as %
\begin{equation}
    U(r,\chi)\>=\>4\epsilon\left[{\left(\frac{\sigma}{r}\right)}^{12}_{}-(1-\chi){\left(\frac{\sigma}{r}\right)}^{6}_{}\right].
    \label{eq:modLJ}
\end{equation}
The information about the relative orientation vectors of the two molecules is embedded in the \( \chi \)-term, connected with the attractive London\citep{London1937} force term. %
For unfavored orientations the attraction is reduced. %
From now one we denote the molecules as particles. %
We chose a cutoff distance for the pair-wise interaction of \(3\,\sigma\), at which the potential energy is approximately \(0\). %
This allows for an exceedingly efficient parallelization of the Monte Carlo algorithm. %

For the definition of the orientation dependent term \( \chi \) we resort to the parameters, characterizing the mutual position of two particles; see figure~\ref{fig:particles_scheme}. %
The optimum configuration is defined via \( \gamma=\beta^{}_1=\beta^{}_2 \), when the orientation vectors of both particles are arranged in a plane. %
The angle \( \gamma \) can be fixed beforehand, thereby determining the energetically optimum size of a micelle. %
Now we seek an expression for \( \chi \) such that for this optimum one has \( \chi=0 \) and otherwise \( \chi > 0 \). %

For the determination of \( \chi \) only the orientation enters. %
The calculation of \( \chi \) is performed by taking both interacting particles and normalize their connection vector to the length of 1. %
The four distances \(d_{\mathrm{a}}\), \(d_{\mathrm{b}}\), \(d_{\mathrm{c_1}}\) and \(d_{\mathrm{c_2}}\), as indicated in figure~\ref{fig:particles_scheme}, can then be directly calculated from the connection vector and the two orientation vectors of the particles. %
In the enthalpic optimum configuration, where \( \gamma=\beta^{}_1=\beta^{}_2 \), these distances are given by %
\begin{equation}
    \begin{aligned}
        d_{\mathrm{a,opt}} &= 1+\kappa \;\mathrm{\sin}(\gamma) &&\\
        d_{\mathrm{b,opt}} &= 1-\kappa \;\mathrm{\sin}(\gamma) &&\\
        d_{\mathrm{c_1,opt}} = d_{\mathrm{c_1,opt}} &= \sqrt{1+\kappa^2_{}\;\mathrm{\cos}^2_{}(\gamma)} && .
    \end{aligned}
\end{equation}
The calculated actual distances in simulations are then compared to this optimum via %
\begin{equation}
    \chi=\sum^{}_{i\in \{\mathrm{a,b,c_1,c_2}\}}{\left(d^{}_{\mathrm{i}}-d_{\mathrm{i,opt}}\right)}^2_{},
\end{equation}
which defines the orientation dependent \( \chi \) term of the potential. %
Note, that \( \chi \) alone is not dependent on the distance of the two particles, but solely on their relative rotational configuration. %
The strength of the orientation dependence, i.e. the magnitude of \(\chi\), can be adjusted via the length of the orientation vector \(\kappa\). %
In what follows we always choose \(\kappa = 1\).

Figure~\ref{fig:potprop} shows the potential energy as a function of distance and \( \chi \). %
For two particles in the optimum configuration \( \chi=0 \) and the resulting potential energy curve is exactly the Lennard-Jones potential. %
As \( \chi \) increases the attractive interaction becomes weaker and the minimum slightly shifts towards higher distances. %
\( \chi\geq1 \) results in exclusively repulsive interactions. %

The potential energy landscapes of two interacting particles offer a more detailed view on the expected properties. %
We fixed the particles in the Lennard-Jones minimum distance and rotated individually with respect to their angles \(\beta^{}_{1,2}\), choosing an optimum value of \(\gamma = 11\grad\). %
This results in the potential energy landscape shown in figure~\ref{sfig:landpp}. %
The plot exhibits a minimum at \( \beta^{}_{1,2}=11\grad \). %
Since we can also think of a second situation where the distances \(a,b\) and \(c\) have the optimum values, which is \( \beta^{}_{1,2}=180\grad - \gamma \), there is also a second symmetry-related potential energy minimum. %
In figure~\ref{sfig:landbr} we took a more detailed look at dependency and synergy of the potential energy as a function of the particles distance \( r \) and their orientation \( \beta \). %
The two minima at \( r/\sigma=1.122 \) and \( \beta=11\grad\wedge169\grad \) follow directly from the very definition of \( \gamma \) as well. %
For \( r/\sigma > 2 \)  the orientation dependence disappears as well. %

Naturally, the ruggedness of the potential energy landscape and thus the expected properties of micellization can be modified by the value of $\kappa$. %
Low values of \( \kappa \) will reduce this dependency until the particles no longer form micelles, but unstructured clusters as one would expect from pure Lennard-Jones particles. %
In the opposite limit very few configurations are energetically favored and the system will be stuck in local energy minima. %
Our present choice $\kappa = \sigma = 1$ is a compromise between both limits. %

\subsection{Simulation}\label{subsec:sim}

Simulations were carried out using the Monte Carlo method with the Metropolis\cite{mc-base,metropolis} algorithm in the canonical (NVT) ensemble. %
In every simulation cycle every particle gets chosen in random order and performs two Monte Carlo steps of which one is a translational and the other a rotational step. %
The step width of a translational step is a random number between \(\pm0.2\,\sigma\) and step width of the corresponding rotational step a random number between \(\pm0.2\,\mathrm{rad}\). %
These values were chosen based on preliminary work and held constant throughout the simulation in order to get comparable time scales at various temperatures. %
They are a compromise between a sufficient number of time steps enabling long time scales and an accurate sampling of the phase space. %
Periodic boundary conditions were applied. %
The times, mentioned in this work, are in units of simulation cycles per particle. %

Equilibration of the surfactant system is a challenge, especially at low temperatures. %
In order to be as insensitive as possible to the initial configuration we start with a random distribution of particles, which are randomly distributed and rotated (as defined in figure~\ref{fig:particles_scheme}) across the cubic simulation box. %
Translation in the box and rotation are handled separately. %
Since micellization behavior and equilibration time are strongly temperature dependent, we varied the temperature range from \(T=0.23\) to \(0.26\) in steps of \(0.01\). %

The size of the simulation box is defined via the number of particles in the system \( N \) and the particle density \( \pinit \). %
We set the number of particles to \( 1000 \) with an optimum micelle size of \( 100 \), which is equal to an optimum angle \( \gamma=11\grad \). %
Note that this optimum is just based on a minimum potential energy. %
We checked for some critical observables, discussed below, that no relevant finite size effects are present. %
The graphs were made using time and ensemble average data from our simulations.

For a few parameter combinations, yielding significant non-equilibrium behavior, we have also performed simulations with simulation boxes, containing four times as many particles. %
In no case we have seen finite size effects. %
Thus, for the analysis of this work, system sizes, comprising 1000 particles, are sufficient. %

\subsection{Excluded volume calculation}\label{subsec:excluded}

\begin{figure}
    \includegraphics{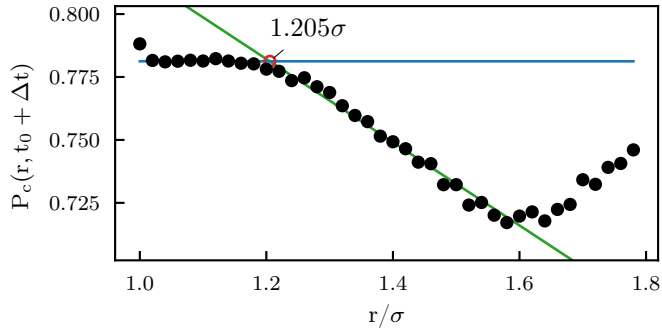}
    \caption{Probability of a pair of particles with a given distance \(r\) at \(t_0\) to be in the same cluster at \(t_0+\Delta t\), \(\pairprobability\) at \(T=0.23\), \(\rho=0.08\) and \(\Delta t=10_{}^4\). %
             Clusters were found using DBSCAN with a threshold of \(1.8\sigma\). %
             For this temperature the critical distance is \(r_\mathrm{crit}^{}=1.205\)}\label{fig:samecluster_r}
\end{figure}

\begin{figure}
    \includegraphics{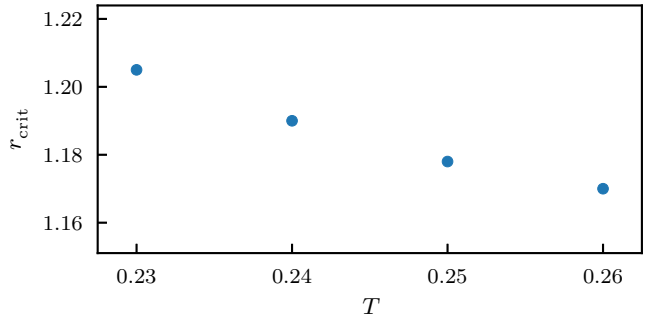}
    \caption{Dependence of \(r_\mathrm{crit}^{}\) on the temperature.}\label{fig:samecluster_temp}
\end{figure}

The concentration of free surfactant molecules in the system cannot be calculated from the number free surfactants and the box volume directly. %
Already aggregated micelles require a fraction of the overall volume and reduce the volume which is accessible to free surfactants. %
As a result, the calculation of the free surfactant particle density must be based on the volume excluded from grown micelles. %
We applied the cluster analysis algorithm density-based spatial clustering of applications with noise\cite{DBSCAN} (DBSCAN) to our system in order to identify particles in cluster structures. %
The DBSCAN algorithm will find clusters in an arbitrary data set along a certain coordinate, which is the distance of particles in our case. %
Since we define a cluster as an agglomeration of particles that stay in spacial proximity for a certain number of time steps, the threshold of the particle distance must be chosen carefully. %
In order to determine the threshold we distinguished between particles in a cluster and nearby particles by taking into account that nearby particles will most likely not stay close to the repulsive outer shell of the micelle for more than a few time steps. %
We therefore calculated all clusters at a given simulation step \(t_0\) with an excessive threshold of \(1.8\sigma\) to find all particles in a cluster and in the close proximity of the cluster. %
At a later simulation step \(t_0+\Delta t\), here \(\Delta t=10^4\), a particle pair of a cluster particle and a particles, which is just passing by, is likely not in the same cluster anymore. %
We observe that the probability of pairs of particles to be in the same cluster \(\pairprobability\), with \(\Delta t=10^4\) time steps, has a plateau value for small particle pair distances at \(t_0\), since these pairs consist of two real cluster particles. %
This probability decreases at a certain particle distance, since these pairs now consist partially of one real cluster particles and one particle passing by. %
We observed this behavior as shown in figure~\ref{fig:samecluster_r} for temperature \(0.23\), where the decrease in probability starts at a pair distance \(r_\mathrm{crit}^{}=1.205\sigma\). %
Interestingly \(r_\mathrm{crit}^{}\) decreases with increasing temperature. %
This may be related to the fact that at fixed distance the binding of close-by particles is no longer as efficient. %
We conclude that in order to accurately calculate the inaccessible volume of an arbitrary cluster we need to run the cluster algorithm with a threshold given by figure~\ref{fig:samecluster_temp} to find all particles in the cluster and exclude particles passing by, which do not belong to the cluster. %

The actual calculation of the inaccessible volume then proceeds as follows. %
We overlay the particle coordinates of a given cluster with a three dimensional grid and an excess of at least \(\pm r_\mathrm{crit}^{}\). %
Every grid point within a distance of \(r_\mathrm{crit}^{}\) of any cluster particle coordinate will be flagged as part of the inaccessible volume. %
The non flagged grid points in the center of the cluster will be flagged as well, so that every flagged point of the grid represents a certain inaccessible volume, which was then calculated as the total of all grid points. %

\subsection{Order parameter calculation}\label{subsec:order_calc}

The spatial realization of micelles can come in all kinds of shapes and structures. %
Since we mainly deal with particles in sphere-like structures we introduce a order parameter to analyze the spatial arrangement of the micelle. %
We start with the order parameter of a given particle \( \orderparticle \) as the scalar product of the normalized orientation vector \( \hat{u} \) and the normalized connecting vector from the centre of mass of the cluster to the considered particle \(\hat{v}\). %
In order to calculate an order parameter of the system \( \ordersystem \), an average of all particle order parameters in clusters of size \( \geq30\% \) of the optimum size of a micelle was calculated.%
\begin{equation}
    \begin{aligned}
        \orderparticle&=\hat{u}\boldsymbol{\cdot}\hat{v}\>\>\in\left[-1;1\right]\\
        \ordersystem&=<\Theta_{p,\Nmin}^{}>\\
    \end{aligned}
\end{equation}
This cutoff for the minimum cluster size was introduced because otherwise the very disordered, small clusters are superimposed to the results of the aggregated micelles which are of key interest. %
The order parameters both can take values between \(\pm1\), where \(1\) represents perfect spherical order and \(-1\) the most disordered state. %

\section{Results}\label{sec:res}

\subsection{Range of applicabilities}\label{subsec:boxes}

\begin{figure}
    \centering
    \subfloat[\label{sfig:micelles}]{ \includegraphics[width=.3\linewidth]{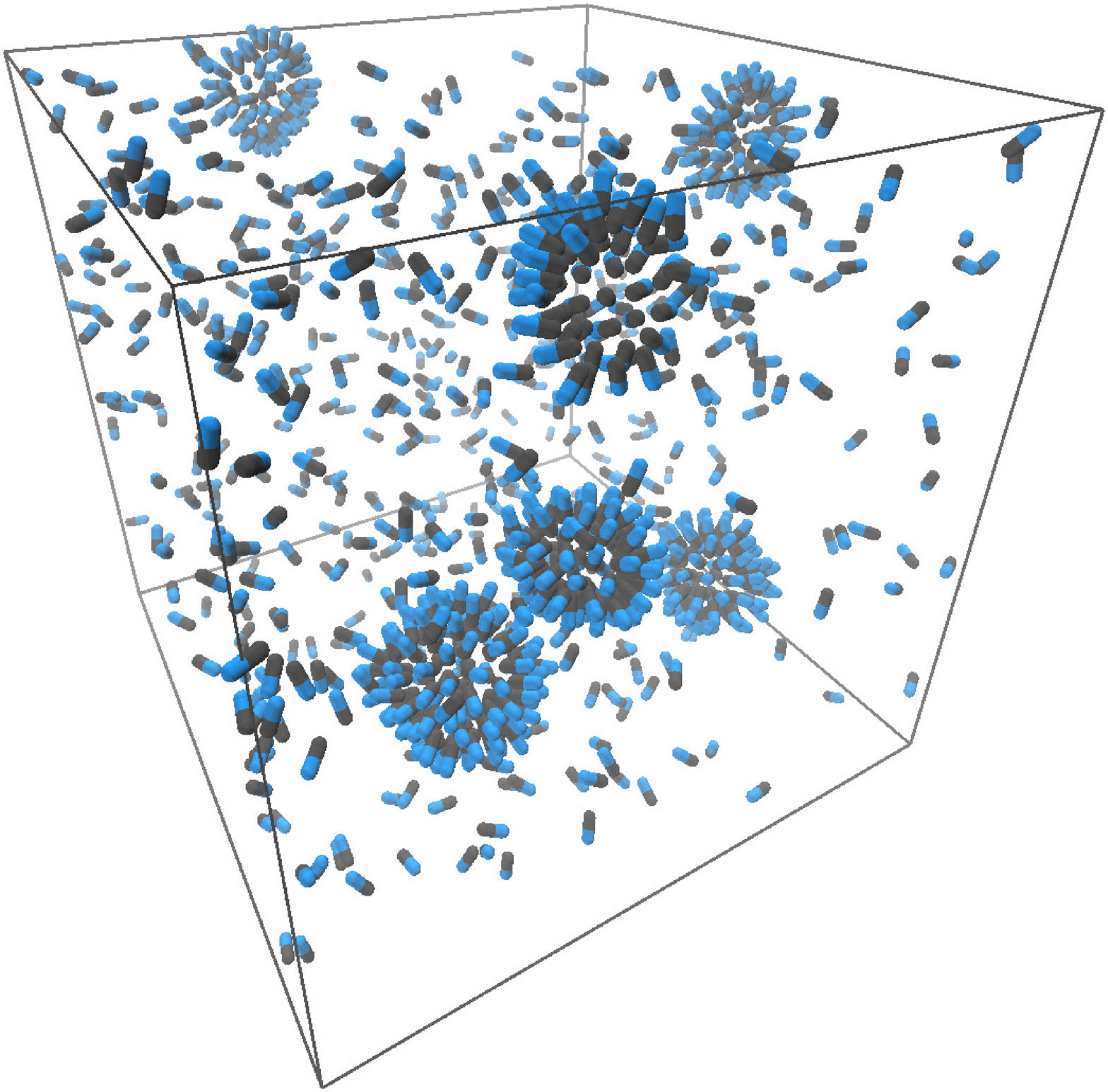} }
    \subfloat[\label{sfig:membrane}]{ \includegraphics[width=.29\linewidth]{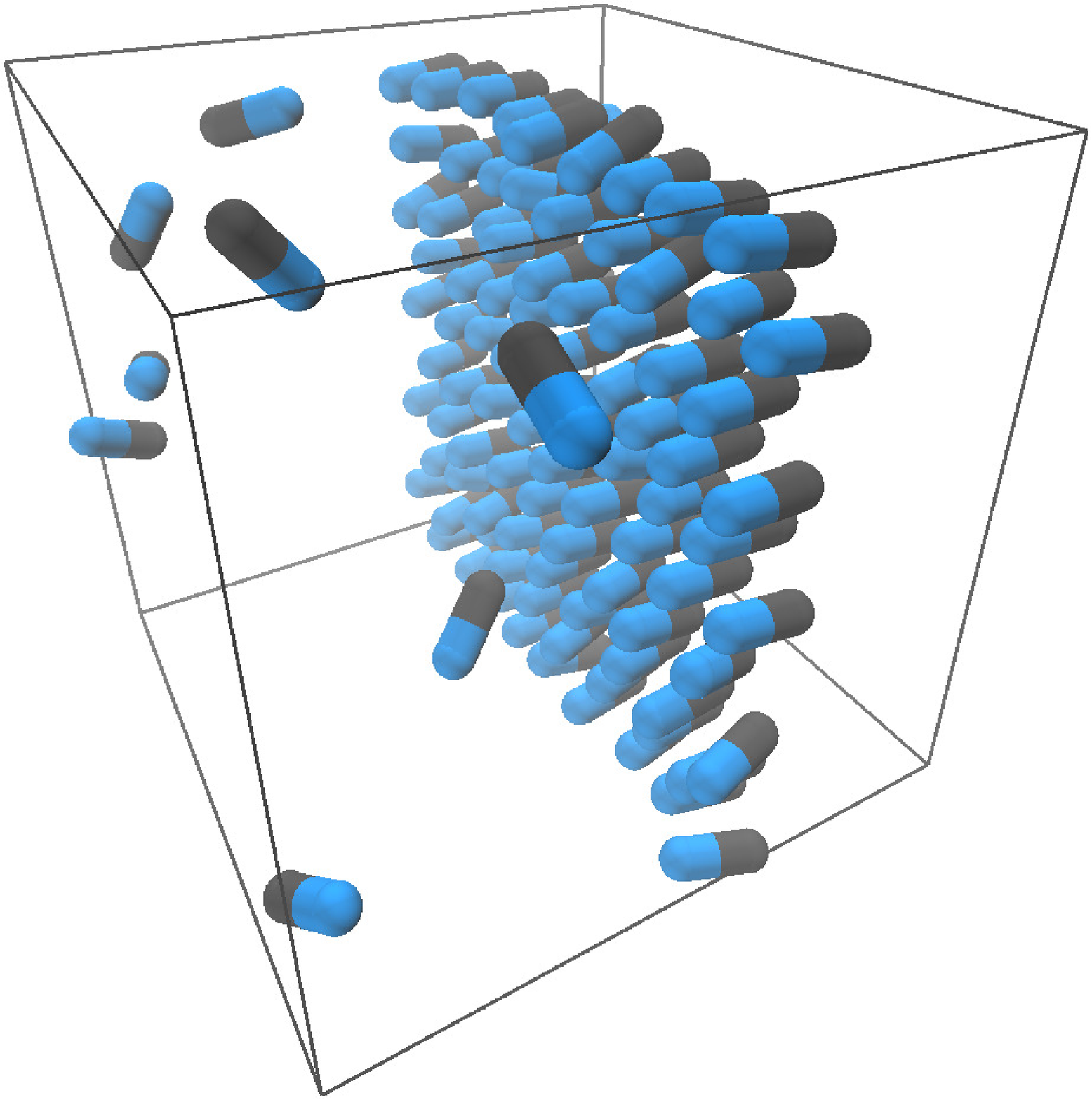} }
    \subfloat[\label{sfig:tube}]{ \includegraphics[width=.3\linewidth]{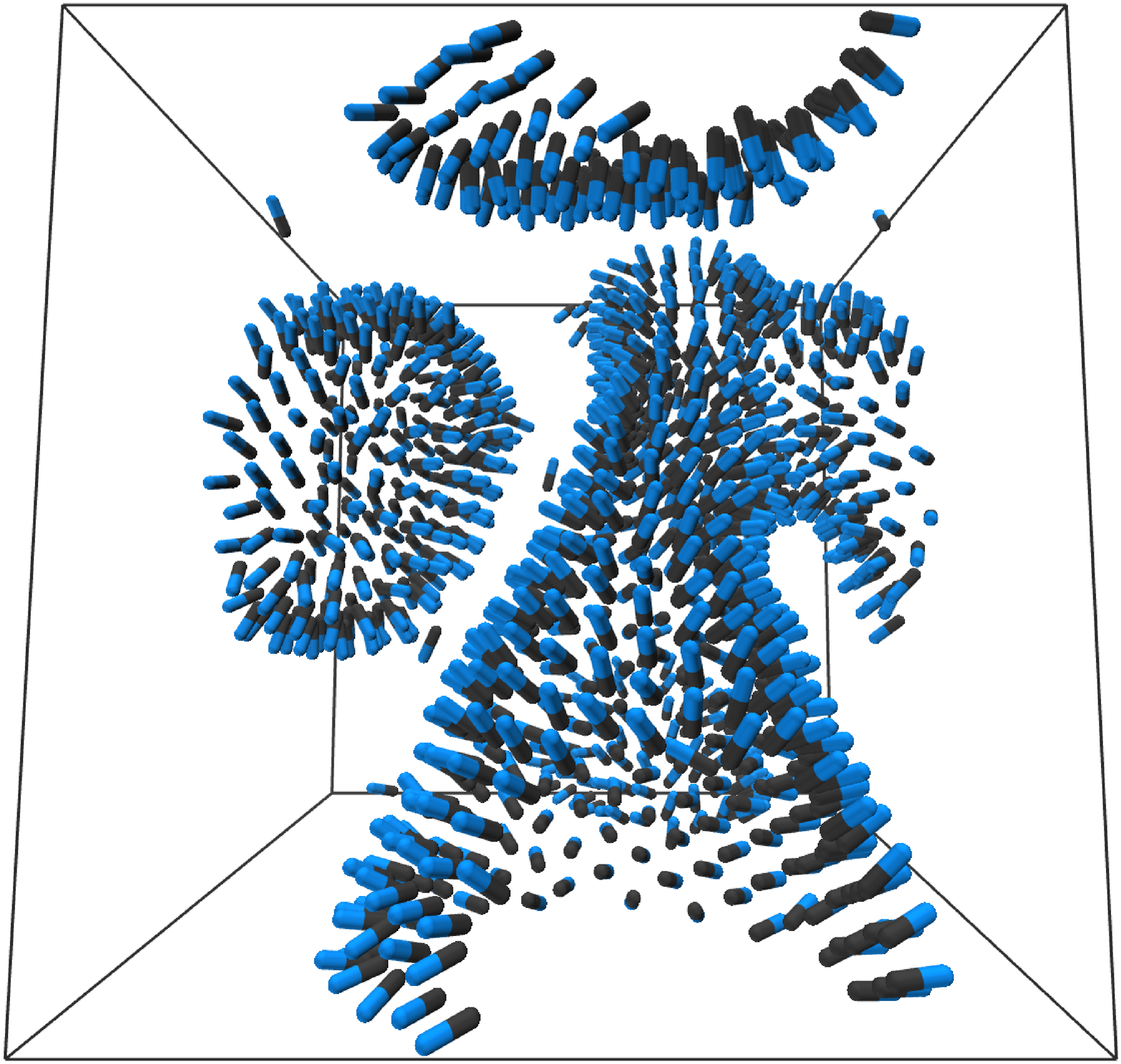} }
    \caption{Simulation screenshots from \refa~a simulation with \( \gamma=11\grad \) forming micelles, \refb~a simulation with \( \gamma=0\grad \) forming a membrane and \refc~a simulation with \(\gamma=5\grad\) at \(T=0.2\) forming a tube. %
    One should note, that in this case a tube like structure clearly is a non-equilibrium effect, where the cluster has not yet resolved into a sphere. %
    Tubes were also observed at higher particle densities and higher temperatures, but were not shown here, since high particle densities result in poor identifiable images. %
    Particles are represented as sticks, where the black and blue part represent the orientation vector \( \hat{u} \), which points in the direction of the blue part. }%
    \label{fig:boxes}
\end{figure}

The behavior of the particles in our model system can be adjusted by various parameters. %
The most obvious influence on the outcome of the system has the choice of the angle \( \gamma \) between two interacting particles which determined the optimum angle between them. %
We show the outcome of two different simulations in figure~\ref{fig:boxes}, where~\ref{sfig:micelles} is a simulation with \( \gamma>0\grad \) and~\ref{sfig:membrane} is a system with \( \gamma=0\grad \). %
A \( \gamma \) value greater than zero clearly results in the formation of micelles in the system, given that the particle density is above cmc. %
The size of the micelle size can be adjusted by the choice of \( \gamma \). %
In this case we chose \( \gamma=11\grad \) which is equivalent to an optimum micelle size of 100 particles. %
Reducing \( \gamma \) will result in larger micelles and even membranes as illustrated in figure~\ref{sfig:membrane}. %
Thus,  in principle our model model can be taken as a starting point for a minimalist modelling of membranes as well. %
However, in the present work we exclusively concentrate on the formation of micelles. %

\subsection{Critical micelle concentration}\label{subsec:cmc}

\begin{figure}
    \includegraphics{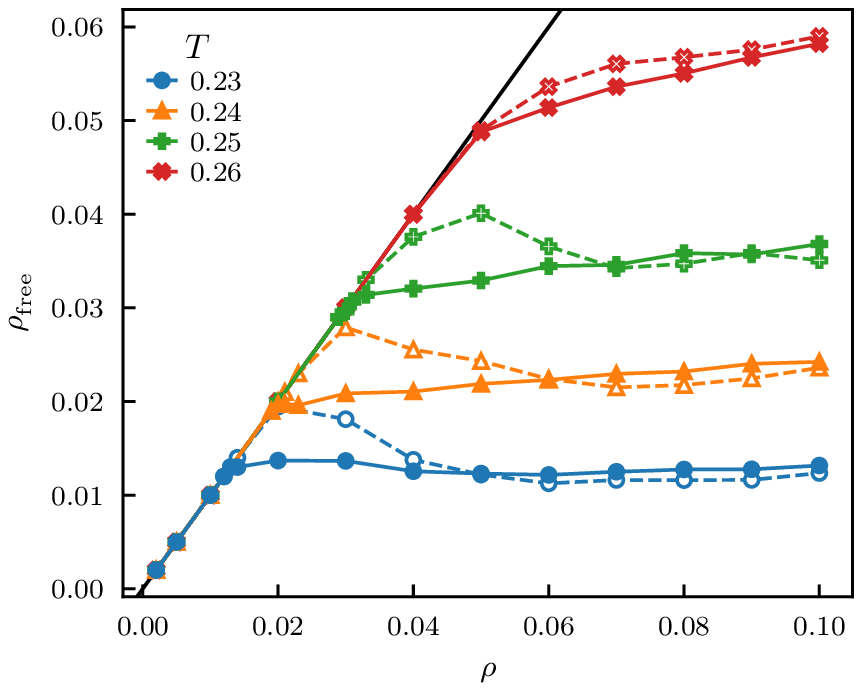}
    \caption{Particle density of free particles \( \pfree \) in the accessible volume for \(1000\) particle systems as a function of the particle density \( \pinit \) at various temperatures of an ensemble and time average from (solid) \( 9.8\cdot10^{7}_{} \) to \( 10^{8}_{} \) and (dashed) \( 3\cdot 10^{4}_{} \) to \( 1.5\cdot 10^{5}_{} \) time steps. %
    The solid black line indicates the linear increase of \(\pfree\). }\label{fig:cfreec}%
\end{figure}

\begin{figure}
    \includegraphics{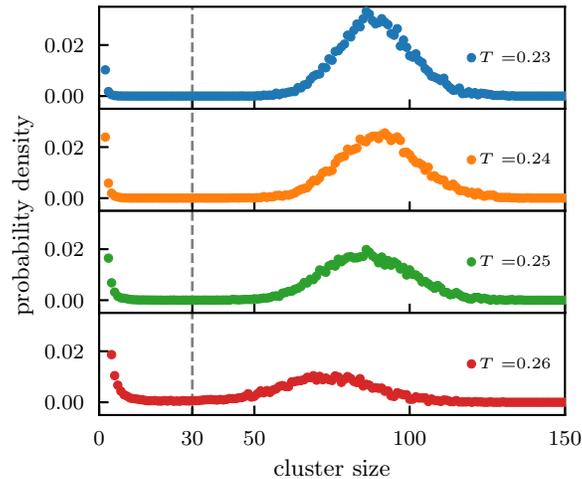}
    \caption{Weighted cluster size distributions of \(\pinit=0.08\) systems at various temperatures of an ensemble and time average from \( 9.8\cdot10^{7}_{} \) to \( 10^{8}_{} \) time steps. %
    Curves are normalized to an area of 1. %
    The dashed grey line at \(N=30\) depicts the threshold at which all systems show loose agglomeration below and micelle formation above.}\label{fig:cluster_size_dist}
\end{figure}

The property of surfactant systems to show micellization behavior is known to be strongly dependent on the concentration of surfactant molecules in the system. %
In a small range of concentrations, surfactant molecules will aggregate as micelles and all additional surfactants will form micelles as well\cite{iupacCMC}. %
The concentration of free surfactants in the system then no longer increases linearly with the overall concentration of surfactants, but remains approximately constant as the overall surfactant concentration increases, as also predicted by theoretical considerations by, \eg, Leibler~\etal\cite{leibler1983theory}. %
This crossover concentration is denoted critical micelle concentration (cmc). %
We calculated the cmc from the data of the particle density of free particles \( \pfree \) as a function of \( \pinit \) (figure~\ref{fig:cfreec}). %
Here, free particles include all particles in clusters smaller than 30 and single particles in the bulk. %
This value was taken as the threshold because cluster size distributions (see figure~\ref{fig:cluster_size_dist}) show a clear differentiation between assembled micelles and loose formations of particles. %
The smaller clusters are formed occasionally even at low concentrations \( \pinit \) for a short period of time and fall apart quickly. %
The increase of \( \pfree \) at low \( \pinit \) therefore is approximately linear. %
Systems with \( \pinit\geq\mathrm{cmc} \) indeed show that \( \pfree \) no longer increases linearly and stays at an approximately constant value. %

\begin{figure}
    \includegraphics{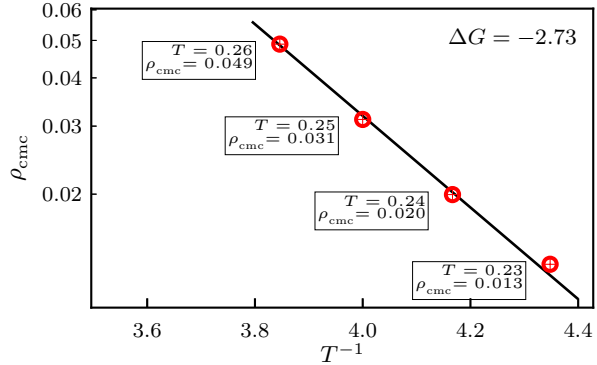}
    \caption{Critical micelle concentration cmc as a function of the inverse temperature \(T^{-1}_{}\) depicted as the red circles. %
    The black line resembles the relation \(\mathrm{cmc}\propto\exp{\Delta G\>T^{-1}_{}}\), where the slope \(\Delta G\) is the Gibbs free energy change of micellization, which is \(-2.73\) in reduced units. }\label{fig:cmc_T_inverse}
\end{figure}

We use the common definition of cmc as the particle density at which micelles start aggregation. %
At higher particle densities, \(\pfree\) approximately reaches a plateau. %
The particles in bulk which are not a member of a any micelle formation then have the highest concentration at which micelle formation is not possible. %
We fit this density domain via linear regression and extrapolate the result towards lower particle densities. %
The intersection of the linear increase of \(\pfree\) at low densities and the fit function is then the critical micelle concentration. 
The value of cmc is shown as a function of the inverse temperature in figure~\ref{fig:cmc_T_inverse}. %

\begin{figure}
    \includegraphics{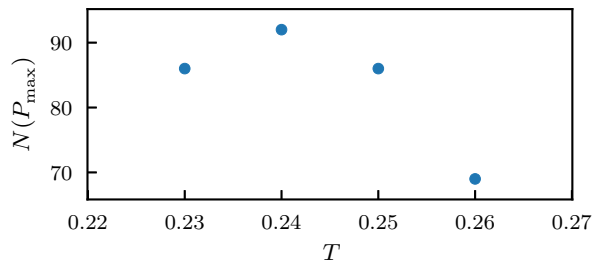}
    \caption{Cluster sizes for clusters \(\geq30\) at maxima of probability densities \(N(P_\mathrm{max}^{})\) from figure~\ref{fig:cluster_size_dist}.}\label{fig:cluster_size_max_N}
\end{figure}

\begin{figure}
    \includegraphics{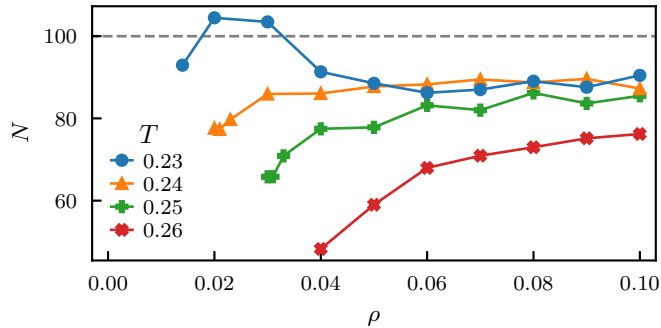}
    \caption{Average cluster sizes for clusters \(\geq30\) as a function of the particle density \(\pinit\) for various temperatures. %
            The dashed grey line depicts the enthalpically optimum size. }\label{fig:N_avg_rho}
\end{figure}

In figure~\ref{fig:cluster_size_max_N} the maxima of the cluster size distributions for cluster sizes larger \(\Nmin\) from figure~\ref{fig:cluster_size_dist} are shown as a function of temperature. %
Interestingly, in particular for higher temperatures the typical cluster size is significantly smaller than the energetically optimum value of \(100\). %
This points towards the relevance of entropic effects favoring the presence of free particles. %
In the low temperature limit, one expects a maximum in cluster size close to \(N_{c}^{}=100\) in case of full equilibrium. %
The system at \(T=0.23\), however, has a maximum in cluster size at \(N_{c}^{}=86\), which is even below the value of \(T=0.24\). %
This non-monotonous behavior suggests the presence of non-equilibration effects even at simulation times as large as \(10^8_{}\). %
Figure~\ref{fig:N_avg_rho} supports the presence of non-equilibrium effects at \(T=0.23\). %
The average cluster sizes in at the two highest temperatures follow a general trend of an increase in cluster size with the increase of \(\pinit\). %
This behavior is expected, since the higher densities result in less available volume for free particles and therefore a reduction in entropy, which favors the enthalpically advantageous clusters. %
However, \(T=0.23\) shows deviations at lower particle densities, where the average size of clusters is generally higher than the trend for higher temperatures would have predicted. %

\begin{figure}
    \includegraphics{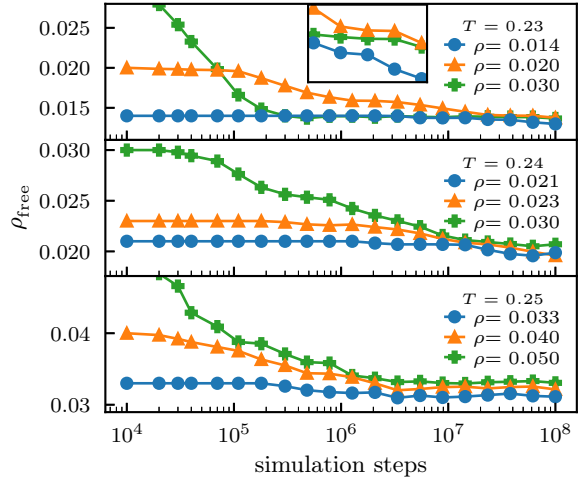}
    \caption{Particle density of free particles \( \pfree \) in the accessible volume as a function of the simulation time at various initial particle densities \( \pinit \) of an ensemble average at various temperatures. The inset in the upper graph for \(T=0.23\) shows the data for \(>10^7_{}\) simulation steps, emphasizing the non-eqilibrium state of the system. }\label{fig:rho_free_time_multi}
\end{figure}

\begin{figure}
    \includegraphics{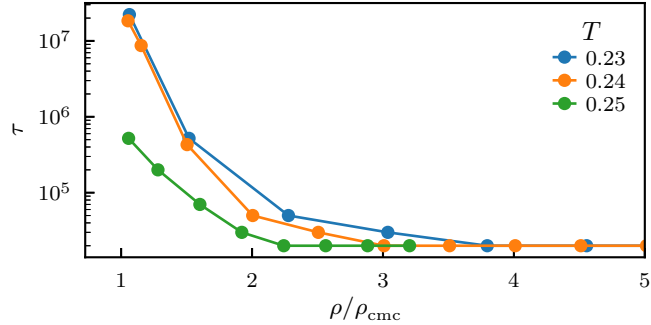}
    \caption{Time scale \( \tau \) to reach the equilibrium number of free particles as a function of particle density normalized to cmc \( \pinit/\pinit_\mathrm{cmc}^{} \) at various temperatures.}\label{fig:tau}
\end{figure}

To elucidate the non-equilibrium effects closer in figure~\ref{fig:cfreec} and~\ref{fig:N_avg_rho} we have also included the dependence of \( \pfree \) on \( \pinit \) in the initial period (\(1.5\cdot10^{5}_{}\) time steps) of the simulation depicted as dashed lines in figure~\ref{fig:cfreec}. %
Obviously, for short times the number of free particles displays a maximum around cmc. %
This indicates the presence of very long equilibration times in that concentration regime. %
This effect is more clearly shown in figure~\ref{fig:rho_free_time_multi} as a function of the simulation time for the relevant particle densities close the cmc domain. %
Temperature \(T=0.23\) is the most significant, since here we found the decrease of \(\pfree\) to be the most pronunced. %
Either way, by taking a closer look at the trend for \(T=0.24\) it is obvious, that also these systems have not reached full equilibrium after \(10^{8}_{}\) simulation steps. %
We state that this behavior fits well to the observation in figure~\ref{fig:N_avg_rho}, where non-equilibrium effects were presumed. %
As an example of systems in full equilibrium in the cmc domain, the curves of \(T=0.25\) show a decrease in \(\pfree\) over time and a convergence to a plateau value. %

In a next step we characterize the time scale to reach equilibrium. %
First we define \(\tau\) as the simulation step, where \(\pfree\) is half way between its maximum in the beginning and its minimum at the end of the simulation. %
This value is shown in figure~\ref{fig:tau}. %
It can be clearly seen that for densities approaching cmc there is a dramatic increase of the equilibration time, in particular for the lower temperatures. %
This resembles the critical slowing down close to phase transitions of second order and clearly shows that very long simulations are required to capture parts of these long-time effects. %
For \(T=0.23\) the relaxation time appears to be larger than \(10^8_{}\) steps. %
In this domain around cmc at low temperatures, \(\tau\) can only show the lower limit of \(\pfree\), since we cannot identify full equilibrium. %
Note that full equilibrium can be even orders of magnitude longer as seen from the slow algebraic time-dependence in the long-time regime in figure~\ref{fig:rho_free_time_multi}. %
As a consequence the system for \(T=0.24\) is still not in equilibrium for \(10^8_{}\) time steps for a broad range of densities although the \(\tau\)-values are significantly smaller than \(10^8_{}\). %
In particular, for long times the value of \(\pfree\) is nearly independent on density \(\rho\), although the \(\tau\)-values display a significant density dependence. %
This may explain why in figure~\ref{fig:cfreec} one still observes a maximum for the lowest temperature and why in figure~\ref{fig:cluster_size_max_N} non-equilibrium effects are observed for the typical cluster size. %
Long equilibration times were also reported for low temperature micellization systems by \citet{cmc-surfactant2} and for diluted systems by \citet{velinova2011sphere}. %
We have excluded the highest temperature \(T=0.26\) from our analysis because equilibration works extremely fast. %
The concentration domain, where long equilibration times occur, is quite narrow and matches the cmc domains in figure~\ref{fig:cfreec}, where \( \pfree \) becomes constant. %

\subsection{Order parameter}\label{subsec:order}

\begin{figure}
    \includegraphics{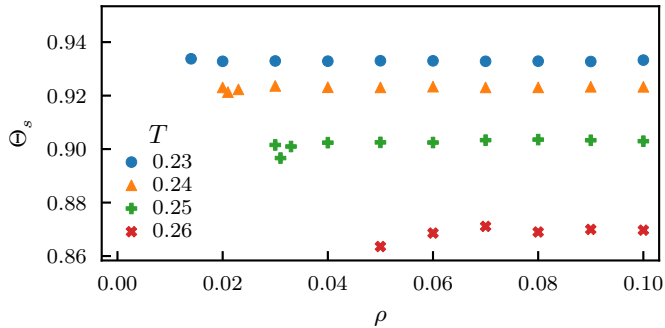}
    \caption{System order parameter \( \ordersystem \) as a function of particle density \( \pinit \) at various temperatures of an ensemble and time average from \( 9.8\cdot10^{7}_{} \) to \( 10^{8}_{} \) time steps for clusters of size \(\Nmin\).}\label{fig:order}
\end{figure}

The system order parameter \( \ordersystem \) (\ref{subsec:order_calc}) is studied in figure~\ref{fig:order} as a function of \( \pinit \). %
By this means, \( \ordersystem \) characterizes the shape of the clusters in the systems with a size \(N_c^{}\geq30\). %
We observed approximately constant values of the order parameter across the whole micellization density region of the systems. %
The different temperatures show a characteristic order value each, which increases with decreasing temperature. %
By comparing the results directly to the trajectories we observed nearly spherical micelles at all temperatures. %
Thus, the particles explore more of their rotational degrees of freedom at higher temperatures. %

\section{Discussion}

We have presented a one bead coarse grained model to perform simulations of micelle forming systems. %
It turns out that an efficient and fast model is required to fully capture the long equilibration time scales, present at lower temperatures and to generate a sufficient number of independent simulation runs, in order to average over the intrinsic fluctuations. %
Furthermore, since we fix the enthalpically optimum micelle size, which would be the equilibrium size in the low-temperature limit, 
we can directly read off the entropic contributions at ambient temperatures. 
Indeed, the temperature dependence of the micelle aggregation numbers, which is characteristic of micelle forming systems \cite{kraft2012modeling,malliaris1985temperature}, is well captured by our model. 
Furthermore also the very slow equilibration at low temperatures close to the cmc domain\cite{velinova2011sphere}, which is a major challenge for computer simulations, is fully recovered by our results (see \ref{subsec:cmc}). %
Due to the long simulation times, possible with our model, we can even quantify the different time regimes relevant for the equilibration process, such as the algebraic time dependence in the long-time aging regime. %
For a reliable quantification of these effects it is also necessary to average over multiple realizations. %
In this way the large fluctuations, related to fast exchange processes of molecules between the environment and a formed micelle, can be averaged out. %

A further key aspect of this work was to develop an algorithm which allows one to accurately estimate the inaccessible volume of the system. %
This is less obvious in continuous space as compared to a lattice model approach. %
We suggested a straightforward approach which can be, as well, applied to different micellar structures such as spheres, ellipsoids and rods. %

One may ask whether it is possible, at least semi-quantitatively, to map a microscopically defined or experimentally measured system on our model system. %
Basically, our model contains the four parameters $\gamma, \epsilon, \sigma$, and $\kappa$. The choice of the equilibrium angle $\gamma$ between two adjacent particles determines the size of the micelles and, thus, is quite straightforward. 

For the choice of the energy scale $\epsilon$ one may refer to our observations in figure~\ref{fig:cmc_T_inverse}.
Interestingly, the relation %
\begin{equation}
    \mathrm{cmc}\propto\exp{\Delta G\>T^{-1}_{}}
\end{equation}
can clearly be observed in agreement, \eg, with the results of simulations of an atomistically resolved model micelle.
The slope can be interpreted as the Gibbs free energy change of micellization with a value of \(\Delta G=-2.73\). %
At the lowest temperature we have a \(\Delta G\>T^{-1}_{}=12.1\). %
This is very close to the value, observed for the micelle formation of DHPC as studied via all-atom simulations (13.9 at the lowest temperature)\cite{kraft2012modeling}. %
Thus, for given temperature this agreement would directly allow one to fix the energy scale \(\epsilon\) so that the values of \(\Delta G\>T^{-1}_{} \) are reproduced.

$\sigma$ is naturally related to the typical equilibrium distance of lipids or surfactants. %
For $\sigma$ one would naturally choose the typical equilibrium distance of lipids or surfactants. %
For this choice, it is instructive to compare \(\pinit\) with the actual particle distance of lipids. %
The typical distance of the centers of mass of two lipids in a bilayer in the first coordination shell is around 1 nm. %
The distance of particles in our system in their first coordination shell is \(\sqrt[6]{2}\approx1.12\) in the enthalpic minimum. %
By taking this conversion factor of \(1\sigma\equiv0.89\mathrm{~nm}\) we find that a particle density of \(\pinit=0.01\) in our system resembles a surfactant solution of \(14.8\textrm{mM}\). %
As a comparison, typically measured critical micelle concentrations of a DHPC system are in the range of 11 to 16 mM solutions at room temperature\cite{kraft2012modeling}, whereas the cmc for our model system approaches values around \(\pinit=0.01\) which is exactly in this regime.%
Thus, despite the minimalistic nature of the model, its behavior resembles the findings for coarse grained and even atomistic simulations in literature\cite{kraft2012modeling}. %

Finally, the model parameter (\(\kappa\)) characterizes the relevance of anisotropic interactions and could, in principle, be related to the nature of the fluctuations of pairs of molecules in a micelle. %
It can, however, not be mapped directly to the length of a lipid molecule, since the inner degrees of freedom of a lipid can not be captured by \(\kappa\) alone. %
A detailed analysis of the impact of $\kappa$ is beyond the scope of this work. %

With the presentation of this model we want to take a step towards the simulation of the frame-guided assembly process proposed by \citet{dong2014frame}. %
As a first step to understand the mechanisms of this process, we propose this robust and versatile one bead model, which is capable of the simulation of simple micelle forming structures. %
This model can easily be extended to model a predefined frame of particles, in which agglomeration can take place with shape and size governed by that frame. %
Furthermore, due to the possibility of performing efficient free energy calculations with this simple model, we envisage to characterize the impact of the frame on the micelle formation (e.g. strong reduction of cmc) in quantitative terms. %
Finally, due to its simplicity we anticipate an efficient scanning of the broad parameter space, which is required to get a profound understanding of frame-guided assembly. %
\setcitestyle{numbers}
Furthermore, a direct comparison with the standard theoretical predictions of micelle formation, \eg[\cite{nagarajan1991theory,moreira2009thermodynamic,santos2016molecular}], would be of major interest. %
\setcitestyle{super}


\bibliography{bibliography}



\end{document}